\documentclass{article}
\usepackage{spconf,amsmath,graphicx,hyperref}
\usepackage{xcolor}
\usepackage{booktabs} 
\usepackage{multirow}
\usepackage{siunitx}
\usepackage{tabularx}
\usepackage{pifont}
\newcommand{\cmark}{\ding{51}} % ✓
\colorlet{wx}{blue}
\title{Dialogue-Based Streaming Audio-Visual Target Speaker Extraction with Predictive Dialogue Information}
\name{
Shuhan Zhang$^{1,2}$,
Wenxuan Wu$^{3}$,
Haizhou Li$^{1,2,*}$\thanks{*Corresponding Author: haizhouli@cuhk.edu.cn}
} 

\address{
$^{1}$ Shenzhen Loop Area Institute, Shenzhen, China \\
$^{2}$ School of Artificial Intelligence, The Chinese University of Hong Kong, Shenzhen, China \\
$^{3}$ The Chinese University of Hong Kong, Shatin, N.T., Hong Kong SAR, China
}
\begin{document}
\ninept
\maketitle

\vspace{-5pt}
\begin{abstract}
In face-to-face, real-time communication, a talking agent must track the
target speaker through natural pauses, turn-taking, and backchannels, often
amid background cross-talk. Most target speaker extraction (TSE) studies,
however, rely on simulated mixtures with full or sparse overlap and ignore
the turn-taking of real conversations. We therefore introduce, to our
knowledge, the first benchmark for online audio-visual TSE (AV-TSE), built
from intact dyadic interactions with independent third-party interference.
Observing that anticipating upcoming activity from semantic, acoustic, and
facial cues benefits online AV-TSE, we propose an LLM-based target-speaker
voice activity projection (TS-VAP) module. Unlike conventional VAP with separated speaker channels, it forecasts the future activity of the target and
conversational partner directly from the overlapping mixture, drawing on the
linguistic and conversational knowledge of a speech-LLM, and uses this
prediction to guide a low-latency separator. We further combine this predictive context with historical and synchronous speaker context. Experiments show that TS-VAP consistently improves streaming extraction across multiple AV-TSE backbones, and that further combining historical, synchronous, and predictive context yields nearly 1 dB gain on real AV conversations.  Project page: \url{https://jjjjiaozi.github.io/TS-VAP/}.

\end{abstract}

\vspace{-5pt}
\section{Introduction}
\label{sec:introduction}

Real-world conversations often unfold in multi-speaker environments, where cross-talk is captured alongside the target speech. Just like humans who listen to one voice at a time, conversational agents are expected to have such attentive ability. 

\begin{figure}[t]
\centering
\includegraphics[width=\columnwidth]
{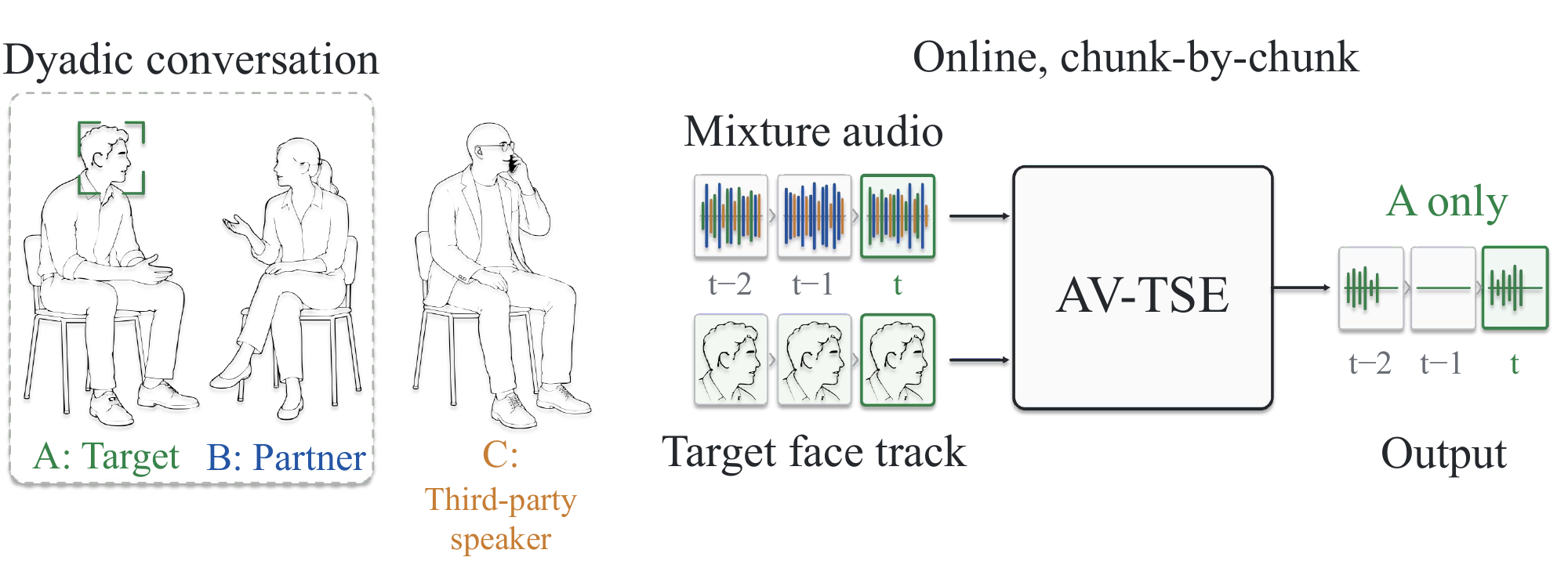}
\vspace{-8mm}
\caption{Online AV-TSE under third-party interference. A and B form a
natural dyadic conversation, while C is an unrelated speaker. Given the
mixture audio and A's face track, the system extracts only A's speech,
suppressing both B and C.}
\vspace{-5mm}
\label{fig:task_overview}
\end{figure}

We consider a natural dyadic conversation with an additional unrelated
third-party speaker, as illustrated in Fig.~\ref{fig:task_overview}. Real
conversations of this kind are increasingly face-to-face and interactive,
as in GPT-4o-style real-time voice assistants~\cite{gpt4o}, where a camera naturally
observes the participant. In such settings the face track is the most
suitable target cue: it fixes the participant's identity without a separate
enrollment utterance, and because it is visual, it remains reliable even
when the acoustic channel is noisy or heavily overlapped. Since the target
must also be recovered as the conversation unfolds, this naturally leads to
online AV-TSE
\cite{zmolikova2023neural,wu2025c,wu2024target, ephrat2018looking,pan2022usev,lin2023avsepformer}.

The central difficulty lies not in the cue but in the conversation itself.
Most online AV-TSE studies evaluate on fully overlapped or manually
simulated sparse-overlap mixtures
\cite{pan2022usev,li2024activeextract,pan2025online}, which combine
utterances independently and lack the turn-taking structure of real
conversations---structure that governs when the target falls silent and
where extraction is hardest. This is especially limiting online: a streaming
model sees no future frames and is thus most ambiguous at turn boundaries
and under similar-sounding, overlapping voices. Predictive knowledge of
upcoming activity would relieve this ambiguity. Voice activity projection
(VAP) can forecast future speaker activity from audio
\cite{ekstedt2022vap,inoue2024realtime,cano2026mmvap,jeon2024llavap}, but is
trained only on clean, single-speaker channels---not the cocktail setting,
where turns must be anticipated from one overlapping mixture, which is
exactly the regime of online AV-TSE.

\begin{figure*}[htbp!]    
\centering    
\includegraphics[width=0.90\textwidth]{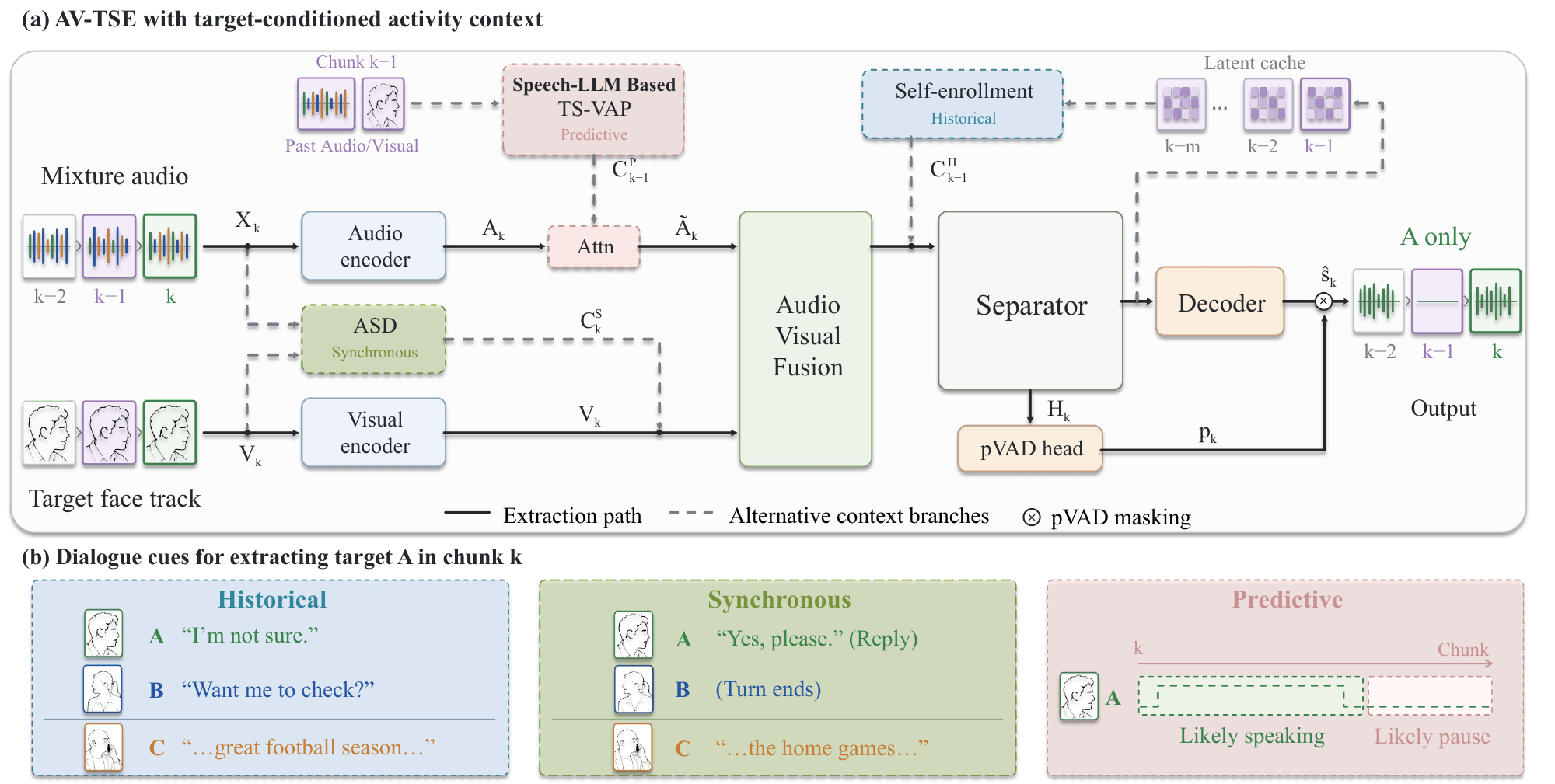}    
\vspace{-3mm}
\caption{(a) AV-TSE extracts target speech in chunk $k$.
Dashed branches denote alternative context schemes:
historical self-enrollment and predictive TS-VAP use information
through $k-1$, while synchronous ASD uses chunk $k$.
TS-VAP anticipates target activity in $k$ and beyond.
(b) Color-matched examples of target--partner dialogue
with unrelated interference; predicted activity is schematic.}
\vspace{-0.3cm}
\label{fig:activity_context}
\end{figure*}

We supply this predictive knowledge through an LLM-based TS-VAP
module. 
% Beyond the clean-channel limitation above, prior VAP models
% anticipate turns mainly from acoustic evidence such as pauses, energy, and
% prosody. What they do not exploit is the linguistic and conversational
% knowledge that governs turn-taking at the level of meaning: a question
% projects an answer, a completed clause invites a hand-off, and a backchannel
% does not. Large language models (LLMs) acquire this knowledge through
% pretraining. Previous studies have shown that text LLMs supply
% transcript-level linguistic knowledge for speech
% extraction~\cite{wu2025incorporating,wu2026elegance}, but have not tapped
% into dialogue-level knowledge. 
Beyond the clean-channel limitation above, conventional VAP models
anticipate turns mainly from acoustic evidence such as pauses, energy, and
prosody. However, linguistic and conversational knowledge that governs
turn-taking at the level of meaning remains underexplored: a question
projects an answer, a completed clause invites a hand-off, and a backchannel
does not. Large language models (LLMs) acquire this knowledge through
pretraining. Previous studies have shown that text LLMs supply
transcript-level linguistic knowledge for speech
extraction~\cite{wu2025incorporating,wu2026elegance}, but have not explicitly
modeled dialogue-level turn-taking knowledge.
We therefore adapt the speech-LLM
Mini-Omni2~\cite{xie2024miniomni2} into a TS-VAP that forecasts
near-future target and interferer activity directly from the overlapping
mixture, anticipating who will speak next from what is being said rather
than from acoustic rhythm alone. We use this forecast to guide a low-latency
separator, and further situate it alongside two other forms of dialogue
context: active speaker detection (ASD) features as synchronous
context~\cite{li2024activeextract}, and self-enrolled speech as historical
context~\cite{li2026memo}.

Our contributions are threefold. \textbf{(1)} We introduce, to our knowledge,
the first benchmark for online AV-TSE on real dyadic dialogues under
third-party interference, built from IEMOCAP~\cite{busso2008iemocap} and a
RealTalk-based test set~\cite{geng2023affective}, whose mixtures retain natural
turn-taking rather than independently assembled utterances. \textbf{(2)} We
propose an LLM-based TS-VAP that transfers a speech-LLM's linguistic and
conversational knowledge into predictive guidance, forecasting future
activity directly from overlapping mixtures. \textbf{(3)} We systematically
compare historical, synchronous, and predictive context for online AV-TSE,
showing where LLM-derived prediction helps across multi-speaker conversation
patterns.

\vspace{-5pt}
\section{Method}
\label{sec:method}

\subsection{Problem Formulation}
Let $s_{\mathrm{tar}}$, $s_{\mathrm{par}}$, and
$s_{\mathrm{ext}}$ denote the target, conversational partner, and
independent third-party speech, respectively. The single-channel mixture is
\begin{equation}
x[n] = s_{\mathrm{tar}}[n] + s_{\mathrm{par}}[n]
+ s_{\mathrm{ext}}[n],
\end{equation}
where $n$ denotes the audio sample index, and the target and partner belong to
the same dyadic dialogue.

For chunk-wise online processing, let $x_k$ and $v_k$ denote the mixture
audio and target face track at step $k$. Target speaker extraction with
predictive context is formulated as
\begin{equation}
\hat{s}_k =
f_{\mathrm{AV\mbox{-}TSE}}
\left(x_k, v_k, C^{\mathrm{P}}_{k-1}\right),
\end{equation}
where $\hat{s}_k$ denotes the reconstructed target waveform at step $k$, and
the predictive context $C^{\mathrm{P}}_{k-1}$ summarizes observations up to
step $k-1$ to predict subsequent speaker activity.

\vspace{-5pt}
\subsection{LLM-based TS-VAP}
We adapt Mini-Omni2 with a Qwen2-0.5B backbone into our
target-speaker-conditioned voice activity projection module, TS-VAP.
% Whisper encodes mixture audio $x$, while CLIP extracts global and
% mouth-region features from target video $v$, augmented with temporal
% differences. Qwen produces visually conditioned audio hidden representations, which are temporally aligned and fused with
% the CLIP-derived visual features, then processed by a lightweight
% causal temporal head:
Whisper and CLIP encode the mixture audio $x$ and target video $v$, respectively. Qwen produces visually conditioned audio representations, which are fused with the CLIP features before a lightweight causal head:

\begin{equation}
U=f_{\mathrm{TS\mbox{-}VAP}}(x,v),
\end{equation}
where $U$ is a sequence of 256-dimensional predictive representations
at 25 Hz, and $U_\ell$ denotes its $\ell$-th frame.
At each frame, $P_\ell=\operatorname{Softmax}(WU_\ell+b)$
predicts joint target--partner activity over the next 2 s,
encoded as $2^8=256$ joint states from four binary future-activity
bins per speaker~\cite{ekstedt2022vap}. Training combines
joint-state classification with auxiliary current-activity and
future-bin supervision. After fine-tuning, TS-VAP is frozen, and
$U$ is used to construct the separator's predictive context.

\begin{table*}[t]
\centering
\caption{
Online extraction results under different backbones and contextual cues.
Context is categorized as historical (History), synchronous (Sync.),
and predictive (Pred.) information, corresponding to self-enrollment,
ASD, and LLM-based TS-VAP prediction, respectively.
RI denotes the relative improvement over the corresponding
context-free baseline,
higher is better.
}
\label{tab:main_results}

\vspace{4pt}
\setlength{\tabcolsep}{4pt}
\renewcommand{\arraystretch}{0.95}

\resizebox{\textwidth}{!}{%
% \begin{tabular}{lccc*{8}{r}}
\begin{tabular}{@{\hspace{1pt}}l@{\hspace{1pt}}ccc*{8}{r}}

\toprule

\multirow{2}{*}{\textbf{Model}}
& \multicolumn{3}{c}{\textbf{Context}}
& \multicolumn{6}{c}{\textbf{TP SI-SNR by overlap ratio (dB)}
$\boldsymbol{\uparrow}$}
& \multirow{2}{*}{\textbf{Avg.} $\boldsymbol{\uparrow}$}
& \multirow{2}{*}{\textbf{RI (\%)} $\boldsymbol{\uparrow}$} \\

\cmidrule(lr){2-4}
\cmidrule(lr){5-10}

& \textbf{History}
& \textbf{Sync.}
& \textbf{Pred.}
& $\boldsymbol{0\%}$
& $\boldsymbol{(0,20]\%}$
& $\boldsymbol{(20,40]\%}$
& $\boldsymbol{(40,60]\%}$
& $\boldsymbol{(60,80]\%}$
& $\boldsymbol{(80,100]\%}$
& & \\

\midrule

% ============================================================
% Backbone baselines
% ============================================================
\multicolumn{12}{c}{\textit{Baselines}} \\[-2pt]
\midrule[0.6pt]

% \multicolumn{12}{l}{\textit{Backbone baselines}} \\[1pt]

Dolphin~\cite{li2026dolphin}$^{\dagger}$
& -- & -- & --
& 2.62 & 2.90 & 2.27 & 3.19 & 3.66 & 4.40
& 3.17 & -- \\

AV-TFGridNet~\cite{pan2023scenario}$^{\dagger}$
& -- & -- & --
& 9.96 & 7.20 & 5.24 & 4.98 & 5.04 & 5.95
& 6.40 & -- \\

TDSE~\cite{wu2019timedomain}
& -- & -- & --
& 11.59 & 10.48 & 6.46 & 5.53 & 5.06 & 5.51
& 7.44 & ref. \\

USEV~\cite{pan2022usev}
& -- & -- & --
& 14.07 & 11.59 & 8.13 & 7.03 & 6.18 & 7.24
& 9.04 & ref. \\

AV-SepFormer~\cite{lin2023avsepformer}
& -- & -- & --
& 15.56 & 11.27 & 7.91 & 6.71 & 6.39 & 6.58
& 9.07 & ref. \\

% \midrule
\midrule[0.6pt]

% ============================================================
% LLM-VAP across different backbones
% ============================================================
\multicolumn{12}{c}{\textit{LLM-VAP across Different Backbones}} \\[-2pt]
\midrule[0.6pt]
TDSE~\cite{wu2019timedomain}
& -- & -- & \cmark
& 13.24 & 11.17 & 6.45 & 5.42 & 4.88 & 5.54
& 7.78 & +4.6 \\

AV-SepFormer~\cite{lin2023avsepformer}
& -- & -- & \cmark
& \textbf{17.25} & 12.06 & 8.29 & 6.70 & 6.48 & 6.73
& 9.59 & +5.7 \\

\multirow{2}{*}{USEV~\cite{pan2022usev}}
& -- & -- & \cmark
& 15.67 & 12.26 & 8.38 & 7.23 & 6.60 & 7.20
& 9.56 & +5.8 \\

% & -- & -- & A
% & 15.24 & 11.98 & 8.25 & 7.19 & 6.47 & 7.35
% & 9.41 & +4.1 \\

& -- & -- & \cmark$^{\mathrm{A}}$
& 15.24 & 11.98 & 8.25 & 7.19 & 6.47 & 7.35
& 9.41 & +4.1 \\

\midrule[0.6pt]
% ============================================================
% Context configurations on USEV
% ============================================================
\multicolumn{12}{c}{\textit{Context Configurations on USEV}} \\[-2pt]
% \midrule
\midrule[0.6pt]

\multirow{6}{*}{USEV~\cite{pan2022usev}}

& \cmark & -- & --
& 16.08 & 12.23 & 8.33 & 7.15 & 6.59 & 7.05
& 9.57 & +5.9 \\

& -- & \cmark & --
& 16.28 & 12.78 & 8.52 & 7.29 & 6.69 & 6.89
& 9.74 & +7.8 \\

& -- & \cmark & \cmark
& 17.09 & 12.75 & 8.17 & 7.02 & 6.41 & 7.17
& 9.77 & +8.1 \\

& \cmark & \cmark & --
& 16.07 & 12.91 & \textbf{8.67} & 7.30 & 6.83 & 7.34
& 9.86 & +9.1 \\

& \cmark & -- & \cmark
& 17.07 & 12.45 & 8.42 & \textbf{7.39} & 6.55 & \textbf{7.38}
& 9.88 & +9.3 \\

& \cmark & \cmark & \cmark
& 17.19 & \textbf{13.25} & 8.52 & 7.27 & \textbf{6.84} & 7.09
& \textbf{10.03} & \textbf{+10.9} \\

\bottomrule
\end{tabular}%
}

\vspace{2pt}

\footnotesize
$\dagger$ Reference backbone without context adaptation. In Pred.,
\cmark{} denotes LLM-VAP and \cmark$^{\mathrm{A}}$ denotes Acoustic VAP.
\vspace{-5mm}

\end{table*}

\subsection{Online AV-TSE with Historical, Synchronous, and Predictive Knowledge}

We evaluate three forms of context for conversational AV-TSE.
\textbf{Historical context}, inspired by
MeMo~\cite{li2026memo}, summarizes past separator latents
with a GRU to obtain $C^{\mathrm{H}}_{k-1}$, which conditions
the AV-fused features through a residual adapter.
\textbf{Synchronous context}, inspired by
ActiveExtract~\cite{li2024activeextract}, combines ASD
embeddings with features from its visual branch to form
$C^{\mathrm{S}}_k$. This cue is concatenated with face
features and projected before AV fusion.
\textbf{Predictive
context} encodes anticipated target--partner activity from past TS-VAP
features, as detailed below.

To construct the predictive context for chunk $k$, we use TS-VAP features $U$ from the last 2 s before
the current chunk, or all available history if shorter. We average the features within
each 0.2-second interval to obtain at most ten tokens:
\begin{equation}
C^{\mathrm{P}}_{k-1}
=\operatorname{Proj}\!\left(
\operatorname{LN}\!\left(
\operatorname{Pool}(U)\right)\right)
+E_{\mathrm{rel}},
\end{equation}
where $\operatorname{Pool}$ is interval-wise mean pooling,
$\operatorname{LN}$ is layer normalization, and
$\operatorname{Proj}$ maps tokens to the audio feature
dimension. $E_{\mathrm{rel}}$ provides learnable positional embeddings
to distinguish recent from older history tokens.

Current audio bottleneck features $A_k$ query this context:
\begin{align}
R_k
&=\operatorname{MHA}\!\left(
\operatorname{LN}(A_k),
C^{\mathrm{P}}_{k-1},C^{\mathrm{P}}_{k-1}\right),\\
\widetilde{A}_k
&=\operatorname{Conv}_{1\times1}\!\left(
[A_k;\operatorname{LN}(R_k)]\right),
\end{align}
where $\operatorname{MHA}$ is four-head attention with
arguments ordered as query, key, and value, and $R_k$
is the retrieved context. Channel concatenation
$[\cdot;\cdot]$ followed by a $1\times1$ convolution
restores the audio feature dimension.
The conditioned features $\widetilde{A}_k$ enter AV fusion;
when no valid context exists, $\widetilde{A}_k=A_k$.

\textbf{Two-pass training}~\cite{li2026memo}.
In the first pass, the separator processes the first $\tau$ seconds
without predictive memory, and TS-VAP features from this prefix
are pooled to construct the memory.
In the second pass, the separator reprocesses the same prefix
together with up to an additional 2 s, using the memory
constructed from the first $\tau$ seconds and keeping it fixed.

% \subsection{Training Objective and Evaluation}

% Prior work handles target-absent mixtures by dividing them into discrete
% activity scenarios with hand-tuned loss weights
% \cite{pan2022usev,li2024activeextract,zeng2025useftp}.
% We found this strategy less effective in our dialogue setting and instead
% add a pVAD head that predicts the frame-level target-activity
% posterior $p_t$, providing explicit supervision over
% target-absent regions~\cite{lin2021sparsely,ding2020personal}. The
% objective is
% \begin{equation}
% \mathcal{L}=\sum_{j=1}^{2}\left(
% \mathcal{L}_{\mathrm{ext}}^{(j)}
% +\lambda_{\mathrm{vad}}\mathcal{L}_{\mathrm{pVAD}}^{(j)}
% \right),
% \end{equation}
% where $\mathcal{L}_{\mathrm{ext}}$ is the negative SI-SNR over
% target-active samples, scaled by their temporal proportion, and
% $\mathcal{L}_{\mathrm{pVAD}}$ is binary cross-entropy. During inference,
% $p_t$ is upsampled to waveform resolution and applied to
% $\hat{s}_t$, the reconstructed target waveform, to suppress
% residual interference in target-absent regions.

\subsection{Training Objective and Evaluation}
Prior work handles target-absent mixtures by dividing them into
discrete activity scenarios with hand-tuned loss weights
\cite{pan2022usev,li2024activeextract,zeng2025useftp}.
We found this strategy less effective in our dialogue setting
and instead add a personal voice activity detection (pVAD)
head~\cite{lin2021sparsely,ding2020personal}, which predicts
the frame-level probability $p_t$ that the target speaker is
active, providing explicit supervision over target-absent regions.
The objective is
\begin{equation}
\mathcal{L}=\sum_{j=1}^{2}\left(
\mathcal{L}_{\mathrm{ext}}^{(j)}
+\mathcal{L}_{\mathrm{pVAD}}^{(j)}
\right),
\end{equation}
where $j$ indexes the two training passes,
$\mathcal{L}_{\mathrm{ext}}$ is the negative SI-SNR over
target-active samples, scaled by their temporal proportion,
and $\mathcal{L}_{\mathrm{pVAD}}$ is binary cross-entropy.
% During inference, $p_t$ is upsampled to waveform resolution
% and multiplied with $\hat{s}_t$, the reconstructed target
% waveform, to suppress residual interference in target-absent regions.
During inference, the target-activity probability \(p_t\) is linearly upsampled to waveform resolution and thresholded to construct an attenuation mask. The reconstructed target waveform is multiplied by this mask, with unit gain in predicted target-active regions and a fixed attenuation factor in predicted target-inactive regions.

% We report SI-SNR for target-present parts and grouped by the overlap ratio
% $r=|\mathcal{A}_{\mathrm{tar}}\cap\mathcal{A}_{\mathrm{int}}|/
% |\mathcal{A}_{\mathrm{tar}}\cup\mathcal{A}_{\mathrm{int}}|$.
We report SI-SNR for target-present parts, grouped by
$r=|\mathcal{A}_{\mathrm{tar}}\cap\mathcal{A}_{\mathrm{int}}|/
|\mathcal{A}_{\mathrm{tar}}\cup\mathcal{A}_{\mathrm{int}}|$,
where $\mathcal{A}_{\mathrm{int}}$ covers partner and third-party activity.
% Prior work handles target-absent mixtures by dividing them
% into discrete activity scenarios with hand-tuned loss weights
% \cite{pan2022usev,li2024activeextract,ao2025used,zeng2025useftp}.
% We found this strategy less effective in our dialogue setting
% and instead add a personal-VAD (pVAD) head that predicts
% frame-level target-activity posteriors $p_k$ from $H_k$,
% supervising both target-active and target-inactive regions
% \cite{lin2021sparsely,ding2020personal}. The objective is
% \begin{equation}
% \mathcal{L}=\sum_{j=1}^{2}\left(
% \mathcal{L}_{\mathrm{ext}}^{(j)}
% +\lambda_{\mathrm{vad}}\mathcal{L}_{\mathrm{pVAD}}^{(j)}
% \right),
% \end{equation}
% where $j$ indexes the two training passes.
% $\mathcal{L}_{\mathrm{ext}}$ is negative SI-SNR over
% target-active samples, scaled by their temporal proportion,
% and is zero for target-absent windows.
% $\mathcal{L}_{\mathrm{pVAD}}$ is binary cross-entropy,
% weighted by $\lambda_{\mathrm{vad}}$.
% During inference, $p_k$ is upsampled to waveform resolution
% and thresholded to gate $\hat{s}_k$, attenuating residual
% interference in predicted target-inactive regions.
\vspace{-5pt}
\section{Experimental Setup}
% \vspace{-3pt}
\label{sec:exp}

\subsection{Datasets}

\noindent\textbf{Vox2Mix pretraining.}
We pretrain all AV-TSE backbones on two-speaker mixtures constructed
from VoxCeleb2~\cite{chung2018voxceleb2}.
Utterances from different speakers are onset-aligned, truncated to
the shorter source, and mixed at a relative level sampled from
$\mathcal{U}(-10,10)$~dB, with mixture duration limited to 6~s.
The training and validation sets contain 20,000 and 5,000 mixtures,
respectively, from 800 speakers.
\noindent\textbf{IEMOCAP-Dialog3Mix.}
We construct 6-s mixtures from IEMOCAP~\cite{busso2008iemocap} by
adding independent third-speaker interference to intact dyadic dialogues,
preserving natural pauses, overlaps, and turn transitions.
The interference duration is sampled from $\mathcal{U}(0,6)$~s and
its level from $\mathcal{U}(-5,0)$~dB relative to the dialogue's pooled
active-speech RMS. Dialogue participants are disjoint across splits.
The training, validation, and test sets contain 23,522, 8,120, and
3,000 mixtures, respectively.

\noindent\textbf{RealTalk-Dialog3Mix.}
To evaluate cross-corpus generalization, we construct an additional
6-s test set from the unscripted, in-the-wild RealTalk
conversations~\cite{geng2023affective}.
We retain annotated dyadic speech in its original timing and add
third-speaker interference following the same procedure as above.
The set contains 1,000 mixtures from 500 dialogue windows, with both
participants used as targets.

\vspace{-5pt}
\subsection{Implementation Details}

We compare five representative AV-TSE extractors spanning different
architectures: Dolphin~\cite{li2026dolphin}, a lightweight audio-visual separation model; AV-TFGridNet~\cite{pan2023scenario}, a
time-frequency GridNet-based extractor; AV-SepFormer~\cite{lin2023avsepformer},
a transformer-based extractor; TDSE~\cite{wu2019timedomain}, a
convolution-based extractor; and USEV~\cite{pan2022usev}, a widely used
dual-path RNN baseline.
All are adapted to the same causal online setting, initialized from
Vox2Mix pretraining, and trained under the same protocol.
We evaluate TS-VAP on AV-SepFormer, TDSE, and USEV to test its consistency across architectures, and run all context combinations on USEV. Every system uses a pVAD head.

% Following~\cite{cano2026mmvap}, TS-VAP is trained separately using 10-s
% observations from the IEMOCAP training split to predict the subsequent 2\,s.
% Its task-specific layers are first trained with Mini-Omni2 frozen at
% $3\times10^{-4}$ for up to 6 epochs. We then jointly optimize these layers and
% rank-8 Qwen LoRA adapters at $10^{-4}$ and $3\times10^{-5}$, respectively,
% for up to 8 epochs. TS-VAP is frozen during AV-TSE training.

% All separators use Adam at $10^{-4}$, a batch size of 32,
% $\lambda_{\mathrm{vad}}=1$, and equal weights for both passes, trained for up
% to 100 epochs with a patience of 10. Online inference uses a 2-s
% initialization/processing window and a 600-ms hop. The pVAD threshold and
% inactive-region attenuation are set to 0.4 and 0.05, respectively, based on
% validation performance.

Following~\cite{cano2026mmvap}, TS-VAP is trained separately using 10-s
observations from the IEMOCAP training split to predict the subsequent 2\,s.
Its task-specific layers are first trained with Mini-Omni2 frozen at
$3\times10^{-4}$ for up to six epochs; we then jointly optimize these layers and
rank-8 Qwen LoRA adapters at $10^{-4}$ and $3\times10^{-5}$, respectively,
for up to eight epochs. TS-VAP remains frozen during AV-TSE training, where all
separators use Adam at $10^{-4}$, a batch size of 32,
and equal weights for both passes, trained for up to 100 epochs with a patience
of ten epochs. Online inference uses a 2-s initialization/processing window and a
600-ms hop, with the pVAD threshold and inactive-region attenuation set to
0.4 and 0.05, respectively, based on validation performance. For comparison, Acoustic VAP uses a TalkNet-based audio-only architecture inspired by~\cite{cano2026mmvap} with the same training protocol as TS-VAP.

\begin{table}[t]
\centering
\footnotesize
\caption{Correct suppression rate (CSR, \%) under non-target interference
when the target is silent. Higher is better.}
\label{tab:vap_suppression}
% \begin{tabular}{@{}lcc@{}}
\begin{tabularx}{\columnwidth}{@{\extracolsep{\fill}}lcc@{}}
\toprule
% \textbf{Model}
\multirow{2}{*}{\textbf{Model}}
& \textbf{Partner}
& \textbf{Third-party} \\
& \textbf{interference}
& \textbf{interference} \\
\midrule
USEV baseline         & 83.96 & 79.73 \\
USEV + Acoustic VAP   & 85.71 & 81.86 \\
USEV + LLM-VAP & \textbf{87.74} & \textbf{85.13} \\
\bottomrule
\end{tabularx}
\vspace{-0.5cm}
\end{table}

\vspace{-5pt}
\section{Results}
\label{sec:results}
\vspace{-5pt}
\subsection{Core Results}
Table~\ref{tab:main_results} summarizes the online extraction results
on 3,000 IEMOCAP-Dialog3Mix test mixtures.
All systems use the same pVAD module and a 2-s processing window
with a 600-ms hop for online inference.
Across the three backbones, TS-VAP consistently improves the average
TP SI-SNR, with relative gains of 5.7\%, 4.6\%, and 5.8\% for
AV-SepFormer, TDSE, and USEV, respectively.
On USEV, TS-VAP achieves 9.56~dB, outperforming Acoustic VAP by
0.15~dB, indicating that target-conditioned audio-visual prediction
provides additional information beyond acoustic prediction alone.
USEV shows the largest relative gain from TS-VAP and is therefore
adopted for the subsequent context-configuration experiments.

% \subsection{Effectiveness of Historical, Synchronous, and Predictive Context}
% We further evaluate historical, synchronous, and predictive context on USEV.
% Individually, the three context types improve the average TP SI-SNR from
% 9.04~dB to 9.57, 9.74, and 9.56~dB, respectively.
% Among their combinations, historical and predictive context achieves
% 9.88~dB, slightly higher than historical and synchronous context at
% 9.86~dB, whereas combining synchronous and predictive context reaches
% 9.77~dB, only marginally above synchronous context alone.
% This suggests that predictive context is more complementary to historical
% speaker information, while partially overlapping with synchronous activity
% cues.
% Using all three further improves the average to 10.03~dB, corresponding to
% a 10.9\% relative improvement over the no-context baseline.
% However, the full configuration does not improve the $(80,100]\,\%$ overlap
% bin (7.09 vs.\ 7.24~dB), suggesting that contextual cues are more useful
% when conversational dynamics and turn transitions provide exploitable
% structure than under near-complete overlap.
\vspace{-5pt}
\subsection{Effectiveness of Historical, Synchronous, and Predictive Context}
We further evaluate historical, synchronous, and predictive context on USEV.
Among the individual cues, synchronous context performs best, while all three
consistently improve over the context-free baseline.
Among pairwise combinations, History+Pred. achieves the best performance,
slightly outperforming History+Sync. and Sync.+Pred., suggesting that predictive
context is more complementary to historical speaker information than to
synchronous activity cues.
Combining all three contexts yields the best overall average of 10.03~dB,
corresponding to a 10.9\% relative improvement over the context-free baseline.
The full configuration, however, provides no gain in the $(80,100]\,\%$
overlap bin, suggesting that contextual cues are most useful when turn-taking
structure remains exploitable.

\vspace{-2mm}

\subsection{Interference Suppression Guided by TS-VAP}

Target-silent regions are challenging for TSE because active non-target speech
can easily leak into the output. This is particularly difficult for unrelated
third-party interference. Predictive context from TS-VAP may help the separator suppress
such leakage by providing cues about the expected activity of the ongoing
conversation.

We evaluate this behavior on all 3,000 IEMOCAP test mixtures under
\emph{partner interference} and \emph{third-party interference}, where the
corresponding non-target speaker is active while the target is silent.
The two conditions may overlap when both interferers are active.
We report the \emph{correct suppression rate} (CSR), measuring how often the
extractor correctly suppresses its output in these target-silent regions.

% As shown in Table~\ref{tab:vap_suppression}, LLM-VAP improves
% CSR over the baseline by 3.78 and 5.40 percentage points under partner and
% third-party interference, respectively. The larger gain for third-party interference suggests that dialogue-aware
% predictive context can also improve robustness to unrelated speech outside
% the target--partner conversational structure.
As shown in Table~\ref{tab:vap_suppression}, Acoustic VAP improves
CSR over the baseline under both interference conditions, while LLM-VAP
provides further gains of 2.03 and 3.27 percentage points for partner
and third-party interference, respectively. Compared with the baseline,
LLM-VAP improves CSR by 3.78 and 5.40 percentage points. The larger gain
under third-party interference suggests that target-conditioned
audio-visual predictive context can improve robustness to unrelated
speech outside the target--partner conversational structure.

% \vspace{-4mm}
% \vspace{-10pt}
\vspace{-5pt}
\subsection{Zero-shot Transfer to RealTalk}

\begin{figure}[t]
    \centering
    \includegraphics[width=0.85\columnwidth]{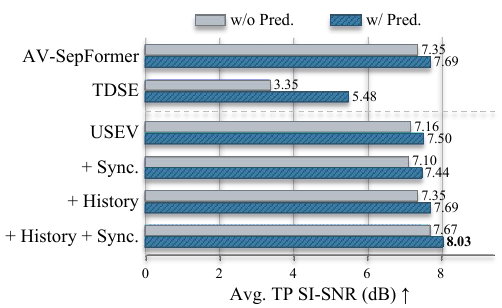}
    \vspace{-0.5cm}
  \caption{Zero-shot online extraction on the 1,000-mixture RealTalk subset.
Each configuration is evaluated with and without the LLM-based TS-VAP. TP SI-SNR in
dB; AV-SepFormer and TDSE report only the baseline, while USEV also includes
sync., history, and their combination.}
    \label{fig:realtalk_results}
    \vspace{-5mm}
\end{figure}

% We further evaluate zero-shot transfer to the RealTalk subset under
% the same online protocol.
% As shown in Fig.~\ref{fig:realtalk_results}, TS-VAP improves all tested
% backbones and context configurations.
% The gain is particularly pronounced for TDSE, from 3.35 to 5.48~dB,
% while AV-SepFormer and USEV improve from 7.35 to 7.69~dB and from
% 7.16 to 7.50~dB, respectively.
% Predictive context also remains beneficial when combined with historical
% and synchronous context, with the full context configuration achieving
% the best result of 8.03~dB.
% These results suggest that TS-VAP remains useful under the RealTalk
% domain shift.
We further evaluate zero-shot transfer to the RealTalk subset under the same online protocol. As shown in Fig.~\ref{fig:realtalk_results}, TS-VAP improves all tested backbones and context configurations. The gain is particularly pronounced for TDSE, increasing from 3.35 to 5.48~dB, while AV-SepFormer and USEV improve from 7.35 to 7.69~dB and from 7.16 to 7.50~dB, respectively. Predictive context also remains beneficial when combined with historical and synchronous context, with the full-context configuration achieving the best result of 8.03~dB. These results demonstrate that the benefits of TS-VAP persist under the RealTalk domain shift.

% \vspace{-10pt}
\section{Conclusion}
We studied streaming AV-TSE in natural conversations with third-party
interference and proposed an LLM-based TS-VAP to provide predictive
conversational context. Results show that predictive context improves
extraction and complements historical and synchronous cues; moreover, the
LLM-based TS-VAP outperforms its acoustic counterpart, and jointly leveraging
all three context types yields the best performance. Future work will integrate the LLM more deeply into the extraction pipeline,
moving from predictive conditioning toward fully semantic-driven separation.

\bibliographystyle{IEEEbib}
\bibliography{strings,refs}

\end{document}